\documentclass[%
onecolumn, 11pt,
superscriptaddress,
nofootinbib,
amsmath,amssymb, amsthm,
aps,
pra,
]{revtex4-2}

\usepackage{graphicx}
\usepackage{dsfont, lipsum}
\usepackage{amsmath,amssymb,amsfonts,amsthm}
\usepackage{enumitem}
\usepackage[table]{xcolor}
\usepackage{mdframed}
\usepackage{centernot}

\usepackage{hyperref}
\hypersetup{
	colorlinks = true,
	linkcolor = blue,
	citecolor = blue,
	urlcolor = blue,
	filecolor = black,
	linktoc = all,
}

\newcommand{\ket}[1]{|{#1}\rangle}
\newcommand{\bra}[1]{\langle{#1}|}
\newcommand{\ketbra}[1]{{\ket{#1} \! \bra{#1} }}

\newcommand{\tr}{\operatorname{tr}}
\newcommand{\one}{\mathds{1}}
\newcommand{\rr}{ {\hspace{1pt} \| \hspace{1pt}} }

\newcommand{\eps}{\epsilon}

\newcommand{\linspan}{{\rm span}}
\newcommand{\bits}{{\rm bits}}
\newcommand{\loge}{\varepsilon}
\newcommand{\Gbar}{\bar{\Gamma}}
\newcommand{\Xbar}{\bar{X}}
\newcommand{\XX}{\mathcal{X}}

\newcommand{\cf}{{\textit{cf.}}}
\newcommand{\ie}{{\textit{i.e.}}}
\newcommand{\eg}{{\textit{e.g.}}}

\newtheorem{thm}{Theorem}
\newtheorem{defn}[thm]{Definition}

\newtheorem{cor}[thm]{Corollary}
\newtheorem{rem}[thm]{Remark}
\newtheorem{prop}[thm]{Proposition}
\newtheorem{lem}[thm]{Lemma}
\newtheorem{ex}[thm]{Example}

\usepackage{pifont}
\newcommand{\circled}[1]{{\large \rm \ding{#1}}}

\definecolor{bc}{cmyk}{0.05,0.05,0,0}
\newmdenv[
    linewidth=0pt,
    backgroundcolor=bc,
    skipabove = 1pt,
    skipbelow = 3pt,
    innerleftmargin=3pt,
    innerrightmargin=2pt,
    innertopmargin=-3pt,
    innerbottommargin=2pt,
]{bbbox}

\newenvironment{proofy}[2]{
{\textit{#1}} {#2} \hfill $\square$
}

\begin{document}

\title{Entropic partial orderings of quantum measurements}

\author{Adam Teixidó-Bonfill}
\email{adam.teixido-bonfill@uwaterloo.ca}
\affiliation{Department of Applied Mathematics, University of Waterloo}
\affiliation{Institute for Quantum Computing, University of Waterloo}
\affiliation{Perimeter Institute for Theoretical Physics, Waterloo}
\author{Joseph Schindler}
\email{josephc.Schindler@uab.cat}
\affiliation{%
F\'{\i}sica Te\`{o}rica: Informaci\'{o} i Fen\`{o}mens Qu\`{a}ntics, Departament de F\'{\i}sica, Universitat Aut\`{o}noma de Barcelona, 08193 Bellaterra, Spain}
\author{Dominik \v{S}afr\'anek}
\email{dsafranekibs@gmail.com}
\affiliation{%
Center for Theoretical Physics of Complex Systems, Institute for Basic Science (IBS), Daejeon 34126, Korea}
\date{\today}  


\begin{abstract}
We investigate four partial orderings on the space of quantum measurements (\textit{i.e.}~on POVMs or positive operator valued measures), describing four notions of coarse/fine-ness of measurement. These are the partial orderings induced by: (1) classical post-processing, (2)~measured relative entropy, (3) observational entropy, and (4) linear relation of POVMs. The orderings form a hierarchy of implication, where \textit{e.g.}~post-processing relation implies all the others. We show that this hierarchy is strict for general POVMs, with examples showing that all four orderings are strictly inequivalent. Restricted to projective measurements, all are equivalent. Finally we show that observational entropy equality $S_M = S_N$ (for all $\rho$) holds if and only if POVMs $M \equiv N$ are post-processing equivalent, which shows that the first three orderings induce identical equivalence classes.
\end{abstract}

\maketitle



\section{Introduction}

In the modern theory of quantum information, any quantum measurement can be mathematically described by a POVM (``positive operator valued measure'', \cf~\cite{davies1970operational,davies1976book,kraus1983book,holevo2011probabilistic,nielsen2010book,wilde2011notes,watrous2018book}) $M$, which is defined (so long as the set of possible outcomes is discrete) as a collection
\begin{equation}
    M = (M_k)_{k \in K},
    \qquad
    M_k \geq 0,
    \qquad
    \textstyle\sum_{k\in K} M_k =\one,
\end{equation}
of positive semidefinite Hermitian operators summing to the identity. Measuring $M$ on a quantum system whose state is given by a density matrix $\rho$, the probability to obtain the outcome labelled by $k$ (which is an element of the outcome set $K$) is given by
\begin{equation}
    p_k = \tr(M_k \rho).
\end{equation}
POVMs include measurements in a complete eigenbasis ($M_k = \ketbra{k}$) and measurements of a Hermitian observable ($M_k$ are projectors into eigenspaces of the observable) as special cases, but more generally can capture the outcome statistics associated with measurements performed by arbitrary sets of quantum operations (\ie~arbitrary quantum instruments~\cite{davies1970operational}). The extension of elementary quantum measurement theory to POVMs and instruments is essential as it allows one to describe \eg~noisy measurements, weak measurements, and the effects of measurement on subsystems~\cite{davies1976book,holevo2011probabilistic,busch2008book,nielsen2010book,wilde2011notes,watrous2018book}.

The set of POVMs has two main mathematical structures that have been studied in detail and thoroughly described: the convex structure~\cite{holevo2011probabilistic,chiribella2004extremal,hein2012foundations,dariano2005classical,chiribella2010barycentric,heinosaari2012coherent,pellonpaa2011complete,sentis2013decomposition,watrous2018book,winter2004extrinsic}, and the post-processing structure~\cite{martens1990nonideal,martens1990inaccuracy,buscemi2005clean,jencova2015comparison,leppajarvi2021postprocessing}. The latter is especially relevant as it describes how the statistics of one quantum measurement can be obtained from another via purely classical operations (indeed merely by manipulation of the outcome labels~\cite{guff2021resource}). This structure takes the form of a coarser/finer partial order on POVMs, where we write (here $N$ is ``coarser'' than $M$)
\begin{equation}
    N \gg M
    \qquad
    \iff
    \qquad
    N_j = \textstyle\sum_i \Lambda_{j|i} \, M_i
\end{equation}
with $\Lambda_{j|i}$ a stochastic map ($\Lambda_{j|i} \geq 0$ and $\sum_{j} \Lambda_{j|i}=1$, \ie~a conditional probability distribution). This stochastic post-processing order induces equivalence classes generated by the equivalence
\begin{equation}
    N \equiv M
    \qquad
    \iff
    \qquad
    N \gg M \gg N.
\end{equation}
Measurements equivalent in this sense can be considered equivalent for all practical purposes, differing only by relabeling and other trivial operations.

The structure of the post-processing partial order was characterized systematically by Martens and de Muynck~\cite{martens1990nonideal}. We recall several of their key findings. One, that each equivalence class defined by $N \equiv M$ has a unique pairwise-linearly-independent representative. Two, that maximally fine (\ie~post-processing ``clean''~\cite{buscemi2005clean,haapasalo2017optimal}) measurements are precisely those of the form \mbox{$M =\big(v_x \ketbra{\varphi_x}\big)_x$} consisting of scalar multiples of rank-1 projectors. And three, that all measurements are finer than $\one$ (the trivial measurement whose only element is the identity), which is the coarsest element. More recently an operational interpretation of the order has been given, first described by Buscemi~\cite{buscemi2018reverse,buscemi2015degradable} (see also~\cite{skrzypczyk2019robustness,lipka2020teleportation,lipka2021nonlocality}), who showed $N \gg M$ holds if and only if~\cite{buscemi2018reverse,buscemi2015degradable,skrzypczyk2019robustness}
\begin{equation}
\label{eqn:guessing-probs}
    P_{\rm guess}(\mathcal{E}|N) \leq P_{\rm guess}(\mathcal{E}|M)
\end{equation}
for all ensembles $\mathcal{E} = (p_u, \rho_u)$, where $P_{\rm guess}(\mathcal{E}|M)$ is probability to guess $u$ correctly after measuring~$M$ on the ensemble. These various properties serve to characterize the set of physically distinct POVMs, and the notions of coarser and finer POVMs, in the sense of classical post-processing.

The post-processing partial order remains a topic of active investigation, and has been studied or applied further in many subsequent works~\cite{martens1990inaccuracy,buscemi2005clean,jencova2015comparison,leppajarvi2021postprocessing,haapasalo2018unavoidable,kimura2010distinguishability,kuramochi2015minimal,dallarno2010purification,dallarno2010purification,wilde2012information,watanabe2012private,farenick2013approximately,oszmaniec2017simulating,guerini2017operational,haapasalo2017optimal,nasser2018characterization,heinosaari2019postprocessing,guff2021resource,heinosaari2021order,lipka2020teleportation,lipka2021nonlocality,buscemi2018reverse,buscemi2015degradable,skrzypczyk2019robustness}. Other related partial ordering structures on quantum measurements and channels have also been considered~\cite{shannon1958ordering,blackwell1953equivalent,winter2001compression,korner1975comparison,buscemi2010comparison,bergmans1973random,gamal1977broadcast,buscemi2014game,torgersen1991comparison,shmaya2005comparison,buscemi2017comparison,gour2018majorization,liu2019resource,skrzpczyk2019all,jencova2021general,hirche2022contraction,gomes2023orders,buscemi2023complete,cohen1998comparisons,lipka2020teleportation,lipka2021nonlocality,buscemi2018reverse,buscemi2015degradable,skrzypczyk2019robustness}.

The post-processing relation $N \gg M$ is a special case of the weaker and more general linear relation
\begin{equation}
    N \Supset M
    \qquad
    \iff
    \qquad
    N_j = \textstyle\sum_i \alpha_{ji} \, M_i,
\end{equation}
where now $\alpha_{ji}$ is an arbitrary (\ie~not necessarily stochastic) real matrix. Relations of this type play an important mathematical role in the theory below, and were also investigated previously~\cite{martens1990nonideal}.

Meanwhile, a crucial role in both classical and quantum information theory is played by entropic quantities such as the Shannon/von Neumann entropies, mutual information quantities, and relative entropies~\cite{shannon1948mathematical,kullback1951information,cover2006book,wilde2011notes}. Here we are particularly interested in the measured entropies of a quantum system, namely the \textit{measured relative entropy} and \textit{observational entropy}, respectively defined by
\begin{equation}
    D_M(\rho \rr \sigma) = \sum_k p_k \log \frac{p_k}{q_k},
    \qquad \qquad
    S_M(\rho) = -\sum_k p_k \log \frac{p_k}{V_k},
\end{equation}
where $p_k = \tr(\rho M_k)$ and $q_k = \tr(\sigma M_k)$ are the probability distributions induced by measuring~$M$ on states $\rho$ and $\sigma$, and $V_k = \tr(M_k)$ are the ``volumes'' of POVM elements. Thus $D_M(\rho \rr \sigma)$ is precisely the classical relative entropy (Kullback--Leibler divergence~\cite{kullback1951information}) between the outcome distributions induced on $\rho,\sigma$, and observational entropy is related to it (in finite dimensions~$d$) by
\begin{equation}
    S_M(\rho) = \log d - D_M(\rho \rr \one/d).
\end{equation}
The observational entropy~\cite{safranek2019a,safranek2019b,safranek2020classical,safranek2021brief,strasberg2020first} and measured relative entropy~\cite{hiai1991proper,vedral1997statistical,hayashi2001asymptotics} have been topics of recent interest in physical~\cite{safranek2019a,safranek2019b,safranek2020classical,safranek2021brief,strasberg2020first,wehrl1979relation,riera2020finite,strasberg2021clausius,amadei2019unitarity,hamazaki2022speed,faiez2020typical,modak2022anderson,stokes2022nonconjugate,strasberg2022book,strasberg2022classicality,safranek2022work,sreeram2022chaos,zhou2023dynamical} and information theoretic~\cite{piani2009relative,buscemi2022observational,zhou2022relations,zhou2022renyi,safranek2023expectation,schindler2020correlation,schindler2023continuity} applications, due to the role of observational entropy connecting coarse-graining to statistical mechanics~\cite{vonNeumann1929proof,vanKampen1954statistics,vonNeumann1955mathematical,wehrl1978general,safranek2019a,safranek2019b,safranek2020classical,safranek2021brief,strasberg2020first}.

The quantity $D_M(\rho \rr \sigma)$ describes how well a measurement $M$ can distinguish between $\rho, \sigma$. If~a~pair $M,N$ are such that $D_N(\rho \rr \sigma) \leq D_M(\rho \rr \sigma)$ for all $\rho,\sigma$, which we denote by
\begin{equation}
\label{eqn:relent-notation}
    D_N \leq D_M,
\end{equation}
then, in the relative entropy sense, $M$ is always a more useful measurement than $N$ for distinguishing between two states. Similarly, $S_M(\rho)$ describes the amount of information missing after a measurement of $M$ on $\rho$ has been performed. If $S_N(\rho) \geq S_M(\rho)$ for all $\rho$, denoted
\begin{equation}
\label{eqn:ent-notation}
    S_N \geq S_M,
\end{equation}
then, in the entropy sense, $M$ always extracts more information than $N$ from the system. We take also analogous notations $D_N = D_M$ and $S_N = S_M$ to denote the cases where (relative) entropy is equal for all (pairs of) states. 

The relations $D_N \leq D_M$ and $S_N \geq S_M$ define two physically motivated ways a measurement $M$ can be finer (more powerful) than another measurement $N$, while the corresponding equalities define two ways that measurements $M,N$ may be equivalent. In this paper we investigate the connection of these two inequality relations both to each other, and also to the stochastic postprocessing relation $N \gg M$. 

Indeed, the motivating questions driving the direction were the following:
\begin{itemize}
    \item Is $S_N = S_M$ equivalent to $D_N = D_M$, and is either equivalent to $N \equiv M$?
    \item Is $S_N \geq S_M$ equivalent to $D_N \leq D_M$, and is either equivalent to $N \gg M$?
\end{itemize}
Both these questions are resolved in the results below. The first is answered in the affirmative, all are equivalent. The second one is answered in the negative, each one is inequivalent to the others. And in the course of resolving the questions, we are led to describe a hierarchy of four partial orders in the space of POVMs.

We remark on connections to other parts of the literature. Comparison of channels, measurements, and statistics, according to various partial orderings, is a topic with a long history~\cite{shannon1958ordering,blackwell1953equivalent} and applications in numerous fields~\cite{cohen1998comparisons}. The most well studied are those related to stochastic post-processings, with order hierarchies dominated by the post-processing (sometimes, ``degradation'') relation studied at least as early as~\cite{korner1975comparison,gamal1977broadcast} for classical channels. In such studies, gaps between the post-processing and other orderings (like we find here) appear to be typical~\cite{korner1975comparison,buscemi2015degradable,buscemi2018reverse}. In the context of POVMs, the stochastic order was elaborated first in~\cite{martens1990nonideal}. In addition to later works directly pertaining to POVMs (\eg~\cite{buscemi2005clean,leppajarvi2021postprocessing,skrzpczyk2019all}), many results on POVM orderings can be viewed as special cases of quantum channel orderings~\cite{buscemi2018reverse,buscemi2015degradable,watanabe2012private,jencova2015comparison,jencova2021general,hirche2022contraction}, where POVMs appear as implemented by quantum-classical channels.

In that channel context a partial ordering closely related to our entropy orderings has been previously studied by Buscemi~\cite{buscemi2010comparison,buscemi2018reverse,buscemi2015degradable}, who studied orders based on mutual information $I(u,x)$ (equivalently conditional entropy $H(u|x)$~\cite{buscemi2018reverse}) between ensembles $\mathcal{E}=(p_u, \rho_u)$ and measurement outcomes $p_x = \tr(M_x \rho)$ of a POVM $M$ acting on $\rho = \sum_u p_u \rho_u$ (\cf~accessible information/Holevo quantity~\cite{nielsen2010book}). That work establishes a gap between the ordering induced by $I(u,x) \leq I(u,y)$ for all ensembles (where $x,y$ label $M,N$ measurement outcomes) and the post-processing order~\cite{buscemi2018reverse,buscemi2015degradable}. We conjecture (but leave open) that the above accessible information ordering is equivalent to the \mbox{$S_N \geq S_M$} ordering appearing in this work. Orderings based on ensemble guessing games have also been of recent interest in the context of distributed/mulitpartite measurements~\cite{lipka2020teleportation,lipka2021nonlocality}.

Another stronger result of Buscemi's work was to show that by replacing conditional $H(u|x)$ with min-conditional entropy $H_{\min}(u|x)$ in the above ordering, the result is identical to post-processing order~\cite{buscemi2018reverse,buscemi2015degradable,skrzypczyk2019robustness} (equivalent to the ``guessing probability'' characterization from \eqref{eqn:guessing-probs} above). This result highlights the need to study the $D_N \leq D_M$ and $S_N \geq S_M$ orderings also in terms of $\alpha$-Renyi divergences. Based on Buscemi's result, one can conjecture that with $D \to D^{\max}$, the post-processing order and our entropic orders will collapse to a single identical one. Closely tied to the guessing probability description of~\cite{buscemi2018reverse,buscemi2015degradable} is the quantification of measurement power by robustness of measurement~\cite{skrzypczyk2019robustness}. Like the quantities appearing in~\cite{buscemi2018reverse,buscemi2015degradable,skrzypczyk2019robustness}, due to our equivalence Theorem~\ref{thm:equality} below, the entropic quantities $D_M(\rho \rr \sigma)$ and $S_M(\rho)$ can each be viewed as a complete family of monotones in the resource theory of quantum measurements~\cite{coeke2016mathematical,skrzypczyk2019robustness,guff2021resource}.

\section{Main results}

In this section we present the main results, and outline the main proofs. Obtaining the difficult parts of the proofs is the topic of the remainder of the paper. The most useful and substantial results are: (1) there exist measurements such that $N$ is entropically coarser than $M$ but $N,M$ are not related by postprocessing (this is strictness of the chain in Theorem 2), and (2) that if $S_N(\rho) = S_M(\rho)$ for all $\rho$ then $N$ and $M$ must be postprocessing equivalent (this is Theorem 4). These are the two nontrivial results whose proofs require the theoretical developments in the rest of the paper. In developing these main theorems a clean structure emerges, in particular a hierarchy of four orderings. We present the results within this hierarchy to give a fuller picture of entropic and postprocessing ordering relations.

We consider a hierarchy of ordering relations for POVMs in which ``finer'' measurements are more informative than ``coarser'' ones in a relevant sense. We restrict to finitely indexed POVMs on finite $d$-dimensional Hilbert space $\mathcal{H}$.

\vspace{6pt}

\begin{bbbox}
\begin{defn}[POVM orderings]
Let $M=(M_i)_i$ and $N=(N_j)_j$ be POVMs. Define the following notions of ordering for POVMs.

\vspace{10pt}

\centerline{
\begin{tabular}{rll}
    \circled{172}
    & 
    $N \gg M$ 
    & 
    (stochastic $N_j = \sum_i \Lambda_{j|i} M_i$) 
    \\[4pt]
    \circled{173}
    & 
    $N \geq M$ 
    & 
    (relative entropy $D_N \leq D_M$) 
    \\[4pt]
    \circled{174}
    & 
    $N \succeq M$ 
    & 
    (entropy $S_N \geq S_M$) 
    \\[4pt]
    \circled{175} 
    & 
    $N \Supset M$ 
    & 
    (linear $N_j = \sum_i \alpha_{ji} M_i$) 
    \\[8pt]
\end{tabular}
}

\noindent
In these definitions $N$ are ``coarser'' and $M$ ``finer'' according to each sense.

\noindent
The notations $D_N \leq D_M$ and $S_N \geq S_M$ have an implied ``for all states'', \cf~\eqref{eqn:relent-notation},\eqref{eqn:ent-notation} above.

\noindent
The stochastic matrices obey $\Lambda_{j|i}\geq 0$, $\sum_j \Lambda_{j|i}=1$, while the linear ones obey $\alpha_{ji} \in \mathbb{R}$.
\end{defn}
\end{bbbox}

These get sequentially weaker in the sense that each one implies the ones below. We find that the hierarchy of the orderings is strict, no two of the relations are equivalent.

\begin{bbbox}
    \begin{thm}
    \label{thm:chain}
    There is a chain of implications
    \circled{172} $\rightarrow$ \circled{173} $\rightarrow$ \circled{174} $\rightarrow$ \circled{175}.\\
    The chain is strict, \circled{172} $\neq$ \circled{173} $\neq$ \circled{174} $\neq$ \circled{175}.
    \end{thm}
\end{bbbox}

\proofy{Outline of proof.}{
The proofs go as follows.
\\
(\circled{172}~$\rightarrow$~\circled{173}) 
Monotonicity of classical relative entropy under stochastic channels~\cite{cover2006book,wilde2011notes}.
\\
(\circled{173}~$\rightarrow$~\circled{174}) 
Entropy $S_M(\rho) = \log d - D_M(\rho \rr \one/d)$ is a special case of relative entropy.
\\
(\circled{174}~$\rightarrow$~\circled{175}) 
If $N \succeq M$ then $S_N(\rho)=\log d$ is implied by $S_M(\rho)=\log d$. But by Prop.~\ref{thm:linearly-related-POVMs} below, this is equivalent to $N \Supset M$.
\\
(Strict) This will be proved by direct construction of counterexamples. For instance, Ex.~\ref{ex:binary-computational-ordering} gives a pair of $N,M$ for which \circled{174} holds but \circled{173} does not. The proof that the entire chain is strict appears as Props.~\ref{thm:separating-1}--\ref{thm:separation-detailed}.
}


Restricted to projective measurements all the orderings are equivalent.
\begin{bbbox}
    \begin{thm}
        For projective measurements 
        \circled{172}=\circled{173}=\circled{174}=\circled{175}.
    \end{thm}
\end{bbbox}

\proofy{Outline of proof.}{
    The idea is that between a pair of projective measurements, the only possible linear relation is a stochastic one, so \circled{175} $\rightarrow$ \circled{172}. The proof is stated as Cor.~\ref{thm:projective-detailed}, which follows from Prop.~\ref{thm:one-projective}.
}

For the stochastic, relative entropy, and entropy orderings, the conditions for equality are equivalent, even though the inequality relations are inequivalent. Thus these three preorderings induce the same equivalence classes.

\begin{bbbox}
    \begin{thm}
    \label{thm:equality}
        The cases of equality for entropy, relative entropy, and stochastic ordering are equivalent,
        $$S_M = S_N \iff D_M = D_N \iff N \equiv M.$$
        The notation $N \equiv M$ indicates post-processing equivalence, meaning $N \gg M \gg N$.
    \end{thm}
\end{bbbox}

\proofy{Outline of proof.}{
    From Thm.~\ref{thm:chain} it is clear that $N \equiv M$ implies $D_N = D_M$ implies $S_N = S_M$ (since two-sided inequality gives equality). The question then is whether $S_N = S_M$ implies $N \equiv M$, completing the implication loop. This is indeed true, and will be shown in Prop.~\ref{thm:equality-detailed}. The fact that this is true is not obvious, because the entropy equality condition $S_M(\rho) = S_N(\rho)$ is difficult to analyze directly.  But the key idea is that if $M$ and $N$ were not stochastically equivalent, then starting from a $\rho_0$ where $S_M(\rho_0)=S_N(\rho_0)$, one would be able to find a way to perturb $\rho$ to make the entropies unequal. This is formalized by a local analysis in terms of derivatives $\frac{d^n}{dt^n}$ of the entropy  $S_M(\rho_t)$ along some parameterized curve $\rho_t$. Using derivatives up to a sufficiently high order, linear algebra arguments show that $M$ and $N$ must have a stochastic linear relation for all the derivatives to be equal. The full proof appears below as Prop.~\ref{thm:equality-detailed}.
}

These theorems characterize the relation and structure between the entropy, stochastic, and linear orderings. Importantly, equivalence between measurements in the entropic senses is the same as the usual stochastic equivalence that has been studied extensively~\cite{martens1990nonideal,martens1990inaccuracy}.

\vspace{6pt}

The remainder of the paper is devoted to proving these main results.

\section{Projective Measurements}

A projective measurement is a POVM $P = (P_x)_x$ such that $P_x P_y = \delta_{xy} P_x$. That is, one whose elements form a complete set of orthogonal projectors, \eg~measuring a Hermitian observable.

If either measurement is projective, some order relations become equivalent.

\begin{bbbox}
    \begin{prop}[One projective measurement] 
    \label{thm:one-projective}
        If $P$ is projective then both
        \begin{equation}
            \begin{array}{ccc}
                N \Supset P & \iff & N \gg P
            \end{array}
        \end{equation}
        and 
        \begin{equation}
            \begin{array}{ccc}
                P \succeq M & \iff & P \gg M .
            \end{array}
        \end{equation}
        A counterexample can be given demonstrating $P \Supset M \centernot\implies P \gg M$.
    \end{prop}
\end{bbbox}

\begin{proof}
    The leftwards implications are trivial from Thm.~\ref{thm:chain}. For the rest:

    ($N \Supset P$)
    Suppose $N_j = \sum_i \alpha_{ji} P_i$. Summing $j$ and resolving identity yields $\textstyle \sum_i \big(1 - \sum_j \alpha_{ji}\big) P_i = 0$, and since $P_i$ are orthogonal it follows that $\textstyle\sum_j\alpha_{ji}=1$. Positivity $\bra{\psi}N_j \ket{\psi} \geq 0$ evaluated on eigenvectors of $P_i$ ensures $\alpha_{ji}\geq 0$.

    ($P \succeq M$)
    Let $\rho=\sum_j c_j P_j$ with distinct $c_j$. Then $S_M(\rho)=S(\rho)$ is minimal so \mbox{$P=M_\rho \gg M$~\cite{safranek2020quantifying}.}

    (Counterexample) On a qubit let $P = \big(\ketbra{0},\ketbra{1}\big)$ and $M=\big(\frac{\ketbra{0}}{2},\frac{\ketbra{1}}{2},\frac{\one}{2}\big)$. Clearly $P \Supset M$. But $\ketbra{0} = aM_0+bM_1+cM_\one$ with $a,b,c\geq 0$ implies $a=2$ and $b,c=0$, so $P \not\gg M$.
\end{proof}

When restricted to pairs of projective measurements, all the orderings are equivalent. 

\begin{bbbox}
    \begin{cor}[Projective measurements]
    \label{thm:projective-detailed}
        For projective measurements $P$ and $P'$,
        \begin{equation}
            P' \Supset P \iff P' \gg P.
        \end{equation}
        Therefore for projective measurements 
        \circled{172}=\circled{173}=\circled{174}=\circled{175}
        are all equivalent.
    \end{cor}
\end{bbbox}

\begin{proof}
    $P' \gg P$ implies $P' \Supset P$ by Thm.~\ref{thm:chain}, and since $P$ is projective also $P' \Supset P$ implies $P' \gg P$ by Prop.~\ref{thm:one-projective}. Thus \circled{172} $\rightarrow$ \circled{173} $\rightarrow$ \circled{174} $\rightarrow$ \circled{175} $\rightarrow$ \circled{172} for projective measurements.
\end{proof}

These properties arise because linear transformations between projective measurements are highly restricted. For general POVMs the situation is more complicated.

\section{Linearly related POVMs}

A key role in the mathematical theory is played by linearly related POVMs.

Consider the linear space $\mathcal{L}_H(\mathcal{H})$ of Hermitian operators in the Hilbert-Schmidt inner product
\begin{equation}
    \langle x,y \rangle = \tr(x^\dag y).
\end{equation}
Denote by $\linspan(M) = \left\lbrace\sum_i c_i M_i \, | \, c_i \in \mathbb{R}, M_i \in M \right\rbrace$ the span in $\mathcal{L}_H(\mathcal{H})$ of elements of $M$. Denote by $\linspan(M)^\perp$ its orthogonal complement, and let $x \perp \linspan(M)$ denote $x \in \linspan(M)^\perp$. We continue to work in finite dimensions $d$ and assume finitely indexed measurements.

Minimal relative entropy (maximum entropy) occurs when $M$ cannot distinguish $\rho$ from $\sigma$. This occurs precisely when the difference $\rho - \sigma$ is orthogonal to all POVM elements of $M$.

\begin{bbbox}
    \begin{prop}[Maximal entropy] \label{thm:linear-POVMs-maximal-entropy}
    The following hold:

    \begin{enumerate}[label=(\alph*), leftmargin=2\parindent]
        \item $D_M(\rho \rr \sigma) = 0$ if and only if $(\rho-\sigma) \perp \linspan(M)$.
        \item $S_M(\rho) = \log d$ if and only if $(\rho-\one/d) \perp \linspan(M)$.
    \end{enumerate}
    \end{prop}
\end{bbbox}

\begin{proof}
    (a) $D_M(\rho \rr \sigma) = 0$ if and only if all $\tr(M_i\rho)=\tr(M_i\sigma)$. Thus $\tr(M_i(\rho-\sigma)) = 0$.

    \noindent
    (b)~Equivalent to $D_M(\rho \rr \one/d)=0$.
\end{proof}

This connects the entropic structure to the linear structure in maximum entropy cases. 

\begin{bbbox}
    \begin{prop}[Linearly related POVMs] \label{thm:linearly-related-POVMs}
    The relation \circled{175}, namely $N \Supset M$, denotes that $N$ is linearly related to $M$ by \mbox{$N_j = \sum_i \alpha_{ji} M_i$}, with $\alpha_{ji}$ a real matrix. The following are equivalent:

    \begin{enumerate}[label=(\alph*), leftmargin=2\parindent, itemsep=0pt]
        \item $N \Supset M$,
        \item $\linspan(N) \; \, \subset \; \linspan(M)$,
        \item $\linspan(N)^\perp \supset \; \linspan(M)^\perp$,
        \item For any $\rho,\sigma$, if $D_M(\rho \rr \sigma) = 0$ then also $D_N(\rho \rr \sigma) = 0$,
        \item For any $\rho$, if $S_M(\rho) = \log d$ then also $S_N(\rho) = \log d$.
    \end{enumerate}
    \end{prop}

    \noindent
    This shows that $N \Supset M$ means $N$ is entropy-coarser than $M$ in maximal entropy cases.
\end{bbbox}

\begin{proof}
    First we prove the chain of forward implications.

    (a) The definition. (b) If $x = \sum_j c_j  N_j$ then also $x=\sum_i (\sum_j c_j \alpha_{ji}) M_i$. (c) If all $\langle x, M_i\rangle = 0$ then by linearity all $\langle x, N_j \rangle=0$. (d) By Prop.~\ref{thm:linear-POVMs-maximal-entropy}, $D_M(\rho \rr \sigma)=0$ implies $(\rho-\sigma)\in \linspan(M)^\perp \subset \linspan(N)^\perp$ which implies $D_N(\rho \rr \sigma)=0$. (e) Equivalent to $D_M(\rho \rr \one/d)=0$.

    Now we prove (e$\to$a). 
    
    We show that if $N \not\Supset M$ then there exists a $\rho$ such that $S_M(\rho) = \log d > S_N(\rho)$.

    Thus if $N \not\Supset M$ then $S_N(\rho) = \log d$ is not implied by $S_M(\rho) = \log d$, a contrapositive proof.
    
    Suppose $N \not\Supset M$. Then there exists an $x$ orthogonal to $\linspan(M)$ but not orthogonal to $\linspan(N)$. Since positive operators form an open set, there exists an $\eps > 0$ such that
    \begin{equation}
        \sigma = \frac{\one + \epsilon x}{d}
    \end{equation}
    is positive. Since $\one \in \linspan(M)$ we also have $\tr(x)=0$ and thus $\sigma$ is a state. We now apply Prop.~\ref{thm:linear-POVMs-maximal-entropy}. Since $\epsilon x/d = (\sigma - \one/d) \perp \linspan(M)$ it follows that $S_M(\sigma)=\log d \geq S_N(\sigma)$. But since $\epsilon x/d = (\sigma - \one/d) \not\perp \linspan(N)$ it also follows that $S_N(\sigma) \neq \log d$. This implies we have found a state such that $S_M(\sigma) = \log d > S_N(\sigma)$.
\end{proof}

In proofs below, POVMs such that both $N \gg M$ and $N \Subset M$ (those related by a linearly invertible stochastic map whose inverse is not stochastic) are found to be particularly useful.

\begin{bbbox}
    \begin{rem} \rm
        POVMs related by $N \gg M$ with also $N \Subset M$ in the reverse direction were studied thoroughly in~\cite{martens1990nonideal,martens1990inaccuracy}. There they interpret a stochastically coarser $N$ as a non-ideal or smeared version of a stochastically finer $M$. They interpret the case where additionally \mbox{$N \Subset M$} as $N$ being an ``invertibly non-ideal'' measurement of the stochastically finer~$M$. And if both $N \gg M$ and $N \ll M$ the two are post-processing equivalent.
    \end{rem}
\end{bbbox}

In summary, the relation $N \Supset M$ indicates $N$ is entropy-coarser in maximal entropy cases.

\section{Disjoint convex combinations}

Mixing measurements with noise will be a useful step in examples below for obtaining highly entropic measurements. We establish in this section some lemmas that will be needed.

Given POVMs $M,M'$, denote by $\lambda M \oplus (1-\lambda) M'$, with $0 \leq \lambda \leq 1$, their disjoint convex combination---that is, their convex combination on the disjoint union of their outcome sets. For example, given $M = (A,B)$ and $M'=(A',B',C')$ are measurements with two and three outcomes, their disjoint convex combination is the five-outcome POVM
\begin{equation}
    \lambda M \oplus (1-\lambda) M' = \Big(\lambda A, \; \lambda B, \;  (1-\lambda) A', \;  (1-\lambda) B', \;  (1-\lambda) C'\Big).
\end{equation}
This corresponds to probabilistically doing one or the other measurement while keeping the outcome labels separate. In a slight abuse of notation we will also denote by $\one$ the trivial POVM whose only element is the identity.

An elementary fact is that measured entropies are linear under disjoint convex combination. 

\begin{bbbox}
\begin{lem}
\label{thm:disjoint-linear}
    Measured relative entropy is linear under disjoint convex combination,
    \begin{equation}
        D_{\lambda M \oplus (1-\lambda)N}(\rho \rr \sigma) = 
        \lambda D_M(\rho \rr \sigma) + (1-\lambda) D_N(\rho \rr \sigma).
    \end{equation}
    As a special case, the same holds for observational entropy.
\end{lem}
\end{bbbox}

\begin{proof}
    Let $Q = \lambda M \oplus (1-\lambda)N$. Then
    $D_{Q}(\rho \rr \sigma) 
    = 
    \sum_{k=1}^{m+n} \tr(Q_k \rho) \log \big(\tr(Q_k \rho)/\tr(Q_k \sigma)\big)
    = 
    \sum_{i=1}^{m} \tr(\lambda M_i \rho) \log \big(\tr(\lambda M_i \rho)/\tr(\lambda M_i \sigma)\big)
    +
    \sum_{j=1}^{n} \tr((1-\lambda)N_j \rho) \log \big(\tr((1-\lambda)N_j \rho)/\tr((1-\lambda)N_j \sigma)\big)
    = 
    \lambda  \sum_{i=1}^{m} \tr(M_i \rho) \log \big(\tr(M_i \rho)/\tr( M_i \sigma)\big)
    +
    (1-\lambda) \sum_{j=1}^{n} \tr(N_j \rho) \log \big(\tr(N_j \rho)/\tr(N_j \sigma)\big)
    $, which is the desired result.
\end{proof}

Another key fact is that disjoint combination with $\one$ preserves post-processing relation. Note that we say $M$ is linearly independent when the elements $M_i$ form a linearly independent set.

\begin{bbbox}
    \begin{lem}
    \label{thm:a-coarseness-lemma}
        Let $N' = (1-\lambda) N \oplus  \lambda \one$.
        Assume $M$ is linearly independent. Then $N' \gg M$ if and only if $N \gg M$.
    \end{lem}
\end{bbbox}

\begin{proof}
    The proofs are as follows.
    
    ($\Leftarrow$) $N' \gg N$ by the map $\Lambda_{j'|j} = (1-\lambda)\delta_{jj'}$ with $\Lambda_{\one|j} = \lambda \delta_{\one i}$, and $\gg$ is transitive.

    ($\Rightarrow$) Suppose $N'_k = \sum_i \Lambda_{k|i} M_i$. This means that $(1-\lambda) N_j = \sum_i \Lambda_{j|i} M_i$ and $\lambda \one = \sum_i \Lambda_{\one|i} M_i$. Thus $N$ and $M$ are linearly related by $ N_j = \sum_i \alpha_{ji} M_i$ with $\alpha_{ji} = \Lambda_{j|i}/(1-\lambda)$. Clearly $\alpha_{ji}\geq 0$. But using $\textstyle\sum_j N_j = \one = \textstyle\sum_i M_i$ it follows that $\sum_i \big(1 - \textstyle\sum_j \alpha_{ji}\big) M_i = 0$. If the elements of $M$ are linearly independent this implies $\sum_j \alpha_{ji}=1$ and thus $\alpha_{ji}$ is stochastic.
\end{proof}

\section{Separating the orderings}
\label{sec:separating}

In this section we show that none of the orderings are equivalent to one another. 

The idea is that one starts with some $M$, and wants to find an $N'$ that is more entropic than it but not stochastically coarser. To do so one first chooses some $N$ that is linear but not stochastic in $M$. Mixing a tiny amount $\lambda$ of this $N$ with a completely ignorant (identity) measurement, one obtains a nearly-completely-ignorant measurement $N'=N_\lambda$ that remains non-stochastic in $M$. By choosing $\lambda$ small enough one can make $N_\lambda$ very ignorant (and thus very entropic), so that $N_\lambda$ is able to provide a measurement more entropic and stochastically unrelated to $M$. Assuming linear independence of (the elements of) $M$ ensures the linear relation between the POVMs is unique, making it easier to show that no stochastic linear relation can exist.

\begin{bbbox}
    \begin{prop}
    \label{thm:separating-1}
    Suppose you are given a pair of POVMs such that:
    \begin{itemize}[topsep=8pt, itemsep=2pt]
        \item $M,N$ are POVMs,
        \item $M$ is linearly independent,
        \item $N \Supset M$, by a linear relation $ N_j = \sum_i \alpha_{ji} \, M_i$ with finite norm $\| \alpha \| \equiv \sum_{ij} |\alpha_{ji}|$,
        \item $N\not\gg M$,
        \item all $N_j \geq \beta \geq 0$ and all $\tr(N_j) \geq v \geq 0$.
    \end{itemize}
    Define for all $\lambda \in [0,1]$ the POVM
    \begin{equation}
        N_{\lambda} = \lambda N \oplus (1-\lambda) \one,
    \end{equation}
    and let
    \begin{equation}
                \lambda' = \frac{\beta}{2 \|\alpha\|^2} ,
                \qquad \qquad
                \lambda'' = \frac{v/d}{2 \|\alpha\|^2}.
    \end{equation}
    \clearpage
    Then the following hold:    
    \begin{itemize}
        \item 
            If $\beta > 0$ then $N_{\lambda'}\,$ is a POVM such that $D_{N_{\lambda'}} \leq D_{M}$ but $N_{\lambda'}\, \not \gg M$.
        \item 
            If $v > 0$ then $N_{\lambda''}$ is a POVM such that $S_{N_{\lambda''}} \geq S_{M}$ but $N_{\lambda''} \not \gg M$.
    \end{itemize}
    Since examples of such $M,N$ are given below, this quickly shows \circled{172}$\centernot \longleftarrow $\circled{173} and \circled{172}$\centernot \longleftarrow $\circled{174}.
    \end{prop}
\end{bbbox}

\begin{proof}
First note that if $\lambda > 0$ then $N_\lambda \not \gg M$ by Lemma~\ref{thm:a-coarseness-lemma}, since $N \not \gg M$ and $M$ is linearly independent. Thus $\beta >0$ implies $N_{\lambda'} \not\gg M$, and likewise $v >0$ implies $N_{\lambda''} \not\gg M$. (At the end of the proof we also show that $\lambda',\lambda''<1$ always, ensuring valid convex coefficients.)

It remains to show the entropy inequalities. To do so, we first observe that
\begin{equation}
    D_{N_\lambda} = \lambda D_N
\end{equation}
by linearity under disjoint convex combinations (\cf~Lemma~\ref{thm:disjoint-linear}), since $D_\one$ is always zero. We then make use of a two-sided Pinsker-type inequality bounding relative entropies in terms of trace distances. In particular we make use of the bound (\cf~(6) and (17) of~\cite{audenart2005continuity}) 
\begin{equation}
\label{eqn:relative-entropy-trace-distance}
    \frac{4 \, \loge \, t^2}{\nu_{\rm min}} \geq D(\mu \rr \nu) \geq 2 \, \loge \, t^2,
\end{equation}
where $\nu_{\rm min}$ is the minimum eigenvalue of $\nu$, and $\loge = \log e$ sets units of information, and $t=\frac{1}{2}\|\mu-\nu\|_1$ is trace distance between $\mu$ and $\nu$ (in our case $\mu,\nu$ will be classical states representing measured probability distributions). Such a bound can be used to relate $M$ and $N$ because of the linear relation $N \Supset M$; due to this linear relationship, the trace norm between the induced probability distributions is limited. In particular, for states $\rho,\sigma$ let
\begin{equation}
    p_i^M = \tr(M_i \rho),
    \qquad
    q_i^M=\tr(M_i \sigma),
    \qquad
    t_M = \tfrac{1}{2}\| p^M - q^M \|_1 ,
\end{equation}
and likewise for $N$. By triangle inequality on the absolute value defining the trace distance, and the definition of $\| \alpha \|$, it follows that
\begin{equation}
    t_N \leq \|\alpha\| \, t_M.
\end{equation}
This is sufficient to relate the entropies using the inequalities stated above. Making use of these observations, we now show for all $\beta,v \geq 0$ and all states $\rho,\sigma$ that
\begin{equation}
    D_{N_{\lambda'}}(\rho \rr \sigma) \leq D_{M}(\rho \rr \sigma),
    \qquad \qquad
    D_{N_{\lambda''}}(\rho \rr \tfrac{\one}{d}) \leq D_{M}(\rho \rr \tfrac{\one}{d}).    
\end{equation}
The latter is equivalent to $S_{N_{\lambda''}}(\rho) \geq S_{M}(\rho)$. Thus, showing these inequalities will complete the proof. Note that while the inequalities hold even when $\beta,v=0$, in that case also $N_\lambda \gg M$ and so no counterexample for separation is provided.

For arbitrary $\sigma$, if $\beta >0$ we have the restriction
\begin{equation}
    q_i^N = \tr(N_j \sigma) \geq \beta >0.
\end{equation}
Meanwhile, for $\sigma = \one/d$, if $v>0$ there is always the restriction
\begin{equation}
    q_i^N = \tr(N_j \one/d) \geq v/d >0.
\end{equation}
Treating both together we have $q_i^N \geq \gamma$ with $\gamma=\beta$ or $\gamma=v/d$ as appropriate.

Uniting this with the above observations, and treating both cases together by the assumption $q_i^N = \tr(N_j \sigma) \geq \gamma >0$ (and plugging in the corresponding $\lambda,\sigma$ values), we finally have
\begin{equation}
    D_{N_\lambda} = \lambda D_N 
    \leq \lambda \frac{4 \, \loge \, t_N^2}{\gamma}
    \leq \lambda \frac{4 \, \loge \, \|\alpha\|^2 t_M^2}{\gamma}
    \leq \lambda \frac{2  \|\alpha\|^2}{\gamma} D_M 
    =  D_M.
\end{equation}
The first equality is disjoint convex linearity. Next comes the upper bound from \eqref{eqn:relative-entropy-trace-distance}, applied to the classical probability distributions. Then the bound on trace distances due to the linear relation. Then lower bound from \eqref{eqn:relative-entropy-trace-distance}, \ie~the Pinsker inequality. And finally the definition of $\lambda$ is applied. This completes the proof.

We return to note that $\lambda',\lambda'' \leq 1/2$. This is shown as follows. First, since $N_j \leq \one$ it follows that $\beta \leq 1$ and $v/d \leq 1$. Then, from $\sum_j \tr(N_j) = d$ we have (with $v_i = \tr(M_i) \in [0,d]$) that 
$1 = \sum_{ji} \alpha_{ji} (v_i/d) \leq \sum_{ji}  |\alpha_{ji}| |v_i/d| \leq \sum_{ji}  |\alpha_{ji}|  = \| \alpha \| $. Thus $\beta/\|\alpha \|^2 \leq 1$ and $(v/d)/\|\alpha \|^2 \leq 1$, so the $N_\lambda$ used are always valid disjoint convex combinations.
\end{proof}

\begin{bbbox}
    \begin{prop}[Separation]
    \label{thm:separation-detailed}
        \circled{172} $\neq$ \circled{173} $\neq$ \circled{174} $\neq$ \circled{175}.
    \end{prop}
\end{bbbox}

\begin{proof}
    (\circled{172} $\neq$ \circled{173}) Example~\ref{ex:binary-computational-ordering-positive}.
    (\circled{173} $\neq$ \circled{174}) Example~\ref{ex:binary-computational-ordering}.
    (\circled{174} $\neq$ \circled{175}) Example~\ref{ex:binary-computational-ordering}.
\end{proof}

\section{Examples and counterexamples}

We give here examples of the type needed for the separation theorem.

One can easily construct a class of examples satisfying the assumptions of Prop.~\ref{thm:separating-1}.

\begin{bbbox}
    \begin{ex} \rm \label{ex:stochastic-with-linear-inverse-general}
    Start with a linearly independent POVM $N$. Hit it with an invertible square stochastic $\Lambda_{i|j}$ to form $M$. Then in terms of the inverse $\alpha_{ji} = (\Lambda^{-1})_{ji}$ one has
    \begin{equation}
        M_i = \textstyle\sum_j \Lambda_{i|j} N_j,
        \qquad \qquad
        N_j = \textstyle\sum_i \alpha_{ji} M_i.
    \end{equation}
    It can easily be checked whether the inverse $\alpha_{ji}$ is stochastic, typically it will not be. The following properties follow immediately since $N$ is linearly independent and $\Lambda$ is invertible:
    \begin{itemize}[itemsep=2pt, topsep=8pt]
        \item $M$ is also linearly independent, and both $\Lambda_{i|j}$ and $\alpha_{ji}$ are unique linear transformations,
        \item $N \ll M$ and $N \Supset M$,
        \item $N \gg M$ if and only if $\alpha_{ji} = (\Lambda^{-1})_{ji}$ is stochastic.
    \end{itemize}
    This suffices to establish examples for Prop.~\ref{thm:separating-1}.
    \end{ex}
\end{bbbox}

\vspace{6pt}
We now specialize this class to increasingly specific cases to be used below.

\begin{bbbox}
    \begin{ex} \rm \label{ex:binary-permutation}
    A specific case of Example~\ref{ex:stochastic-with-linear-inverse-general}. Let a linearly independent binary POVM be
    \begin{equation}
        N = \big(A,B\big).
    \end{equation}
    Define for $0<\eps< 1/2$ a stochastic $\Lambda_{i|j}$  and its inverse $\alpha_{ji} = (\Lambda^{-1})_{ji}$ by
    \begin{equation}
        \Lambda_{i|j} =
        \begin{pmatrix}
            1-\eps & \eps \\
            \eps & 1-\eps
        \end{pmatrix},
        \qquad \qquad
        \alpha_{ji} =
        \frac{1}{1-2\eps} \;
        \begin{pmatrix}
            1-\eps & -\eps \\
            -\eps & 1-\eps
        \end{pmatrix}.
    \end{equation}
    The inverse is never stochastic since it has negative values. (Note $\| \alpha \|=2/(1-2\eps)$.) Then
    \begin{equation}
        M = \Big((1-2\eps)A + \eps \, \one, \; \;
        (1-2\eps)B + \eps \, \one \Big),
    \end{equation}
    in accordance with $M_i = \sum_j \Lambda_{i|j} N_j$ and $N_j = \sum_j \alpha_{ji}M_i$. This is equivalent to mixing $N$ with $\eps$ of its permutation, or to mixing $N$ with $2\eps$ of pure noise, the same as the example used for entropy continuity bounds~\cite{schindler2023continuity}. It is clear that $M$ and $N$ are both linearly independent, and that\\[-6pt]
    \begin{equation}
        \begin{array}{c}
             N \not\gg M, \\[2pt]
             N \Supset M, 
        \end{array}
        \qquad \qquad   
        \begin{array}{c}
             N \ll M, \\[2pt]
             N \Subset M,
        \end{array}
    \end{equation}

    \vspace{-4pt}

    \noindent
    by comparing with the conditions given in Example~\ref{ex:stochastic-with-linear-inverse-general}.
    \end{ex}
\end{bbbox}

In the following example the relation of $M$ to $N$ demonstrates \circled{174} $\centernot \longleftarrow$ \circled{175}, while the relation of $M$ to $N_\lambda$ demonstrates \circled{173} $\centernot \longleftarrow$ \circled{174}. This is achieved using a case of Prop.~\ref{thm:separating-1} where $v>0$ but $\beta = 0$. The relation of $M$ to $N_\lambda$ also demonstrates \circled{172} $\centernot \longleftarrow$ \circled{174}.

\begin{bbbox}
    \begin{ex} \rm \label{ex:binary-computational-ordering}
    Consider Example~\ref{ex:binary-permutation} on a qubit, with $A=\ketbra{0}$ and $B=\ketbra{1}$ and $\eps=1/4$, which gives
    \begin{equation}
        N = \Big( \ketbra{0}, \, \ketbra{1} \Big),
        \qquad 
         M = \Big(\tfrac{1}{2} \ketbra{0} + \tfrac{1}{4} \, \one, \; \;
           \tfrac{1}{2} \ketbra{1} + \tfrac{1}{4} \, \one \Big).
    \end{equation}
    As in Example~\ref{ex:binary-permutation} these obey
    \begin{equation}
        \begin{array}{c}
             N \not\gg M, \\[2pt]
             N \Supset M, 
        \end{array}
        \qquad \qquad   
        \begin{array}{c}
             N \ll M, \\[2pt]
             N \Subset M.
        \end{array}
    \end{equation}
    Furthermore, $\tr(N_j) \geq v$ with $v=1$, and the linear relation has $\| \alpha \| = 4$. Therefore Prop.~\ref{thm:separating-1} shows that
    \begin{equation}
        N_{\lambda} = \lambda N \oplus (1-\lambda) \one,
        \qquad \qquad
        \lambda = \frac{v/d}{2 \|\alpha\|^2} = \frac{1}{64},
    \end{equation}
    is a POVM such that $N_{\lambda} \not\gg M$ and for all $\rho$
    \begin{equation}
        S_{N_{\lambda}}(\rho) \geq S_{M}(\rho).
    \end{equation}
    Thus $N_{\lambda} \succeq M$ but $N_{\lambda} \not\gg M$, which proves \circled{172} $\centernot \longleftarrow$ \circled{174}. This is exhibited in one case by evaluating observational entropies on the state $\rho = \ketbra{0}$, yielding
    \begin{equation}
        S_N(\rho) = 0 \; \bits,
        \qquad \quad
        S_M(\rho) = \tfrac{1}{2} + \tfrac{3}{4} \log \tfrac{4}{3}
        \approx 0.81 \; \bits,
        \qquad \quad
        S_{N_{\lambda}}(\rho) = \tfrac{63}{64}
        \approx 0.98 \; \bits.
    \end{equation}
    This shows that $N \Supset M$ but $N \not \succeq M$, which proves \circled{174} $\centernot \longleftarrow$ \circled{175}. On the other hand consider relative entropies on the states $\rho=\ketbra{0}$ and $\sigma=\ketbra{1}$,
    \begin{equation}
        D_N(\rho \rr \sigma) = \infty,
        \qquad \quad
        D_M(\rho \rr \sigma) = \tfrac{1}{2} \log 3
        \approx 0.79 \; \bits,
        \qquad \quad
        D_{N_{\lambda}}(\rho \rr \sigma) = \infty.
    \end{equation}
    (Note: recall that the defining inequalities $S_N \geq S_M$ and $D_N \leq D_M$ go in opposite directions.)\\[4pt]
    \noindent
    Thus $N_{\lambda} \succeq M$ but $N_{\lambda} \not \geq M$, which proves \circled{173} $\centernot \longleftarrow$ \circled{174}.
    \end{ex}
\end{bbbox}

In the following example, a case of Prop.~\ref{thm:separating-1} with $\beta >0$, the relation of $M$ to $N_\lambda$ demonstrates that \circled{172} $\centernot \longleftarrow$ \circled{173}.

\begin{bbbox}
    \begin{ex} \rm \label{ex:binary-computational-ordering-positive}
    As in Example~\ref{ex:binary-computational-ordering}, again consider Example~\ref{ex:binary-permutation} on a qubit, with $\eps=1/4$, but this time with
    \begin{equation}
        N = \Big(\tfrac{1}{2} \ketbra{0} + \tfrac{1}{4} \, \one, \; \;
           \tfrac{1}{2} \ketbra{1} + \tfrac{1}{4} \, \one \Big),
        \qquad 
        M = \Big(\tfrac{1}{4} \ketbra{0} + \tfrac{3}{8} \, \one, \; \;
           \tfrac{1}{4} \ketbra{1} + \tfrac{3}{8} \, \one \Big).
    \end{equation}
    Again
    \begin{equation}
        \begin{array}{c}
             N \not\gg M, \\[2pt]
             N \Supset M, 
        \end{array}
        \qquad \qquad   
        \begin{array}{c}
             N \ll M, \\[2pt]
             N \Subset M.
        \end{array}
    \end{equation}
    And here $N_j \geq \beta$ with $\beta=1/4$ and the linear relation has $\| \alpha \| = 4$. Therefore Prop.~\ref{thm:separating-1} shows
    \begin{equation}
        N_{\lambda} = \lambda N \oplus (1-\lambda) \one,
        \qquad \qquad
        \lambda = \frac{\beta}{2 \|\alpha\|^2} = \frac{1}{128},
    \end{equation}
    is a POVM such that $N_{\lambda} \not\gg M$ and for all $\rho,\sigma$
    \begin{equation}
        D_{N_{\lambda}}(\rho \rr \sigma) \leq D_{M}(\rho \rr \sigma).
    \end{equation}
    Thus $N_{\lambda} \geq M$ but $N_{\lambda} \not\gg M$, which proves \circled{172} $\centernot \longleftarrow$ \circled{173}. As a check, again consider relative entropies on $\rho=\ketbra{0}$ and $\sigma=\ketbra{1}$,
    \begin{equation}
    \begin{array}{c}
         D_N(\rho \rr \sigma) = \tfrac{1}{2} \log 3
        \approx 0.79 \; \bits,
        \qquad \quad
        D_M(\rho \rr \sigma) = \tfrac{1}{4} \log \tfrac{5}{3}
        \approx 0.18 \; \bits,
        \\[12pt]
        D_{N_{\lambda}}(\rho \rr \sigma) = \tfrac{1}{128} \tfrac{1}{2} \log 3
        \approx 0.006 \; \bits.
    \end{array}
    \end{equation}
    This is consistent with $N_{\lambda} \geq M \gg N$ while $N_{\lambda} \not \gg M$, as was claimed.
    \end{ex}
\end{bbbox}

This establishes all the desired separations between the relations. Evidently the ``binary mixed with permutation'' examples sufficed to show all desired properties.

\section{The case of equality}

Although the inequality relations \circled{172} through \circled{175} are all inequivalent, the corresponding equality relations for \circled{172} through \circled{174} are all equivalent.

\begin{bbbox}
    \begin{prop}[Equality]
    \label{thm:equality-detailed}
        If $S_M = S_N$ then $M \equiv N$. That is, if two measurements induce the same observational entropy on all states, then they are stochastically equivalent.
    \end{prop}
\end{bbbox}

\begin{proof}
    Suppose $S_M(\rho) = S_N(\rho)$ for all $\rho$. The idea of the proof is to observe that $S_M(\rho_t) = S_N(\rho_t)$ for any parameterized $\rho_t$, and therefore the derivatives $\frac{d^n}{dt^n} \big(S_M(\rho_t) - S_N(\rho_t) \big)=0$ must all vanish. For convenience we work with $D_M(\rho_t \rr \tfrac{\one}{d}) = D_N(\rho_t \rr \tfrac{\one}{d})$, the equivalent relative entropy form. We consider finitely indexed POVMs in finite dimensions.

    We will evaluate derivatives about the maximally mixed state at $t=0$, in an arbitrary direction. To do so consider the parameterized $\rho_t = \rho(t)$ defined by
    \begin{equation}
        \rho_t = \tfrac{\one}{d} + t X,
    \end{equation}
    where $X$ is an arbitrary traceless Hermitian operator. This is a state for all $t$ in some interval $t \in (-\eps,\eps)$ since positive definite operators are an open set. Denote the derivative of relative entropy along this curve at $t=0$ by
    \begin{equation}
        z_n = \frac{d^{n}}{dt^{n}} \Big[ D_M(\rho_t \rr \tfrac{\one}{d}) \Big] \Big|_{t=0}.
    \end{equation}
    For the $n=0$ case and the $n=1$ derivative we have $z_n=0$, as expected since $\rho_0=\one/d$ is a local minimum with a value of zero. The remaining derivatives $n \geq 2$ yield
    \begin{equation}
        z_n = (-1)^{n} \, (n-2)! \; d^{n-1}  \,  \sum_i \tr(M_i) \, (m_i)^{n},
        \qquad \qquad
        m_i = \frac{\tr(M_i X)}{\tr(M_i)}.
    \end{equation}
    These $n\geq 2$ derivatives will be the ones constrained in the main proof, to which we now turn.

    The equality $S_M(\rho_t)=S_N(\rho_t)$, expressed in the relative entropy form from earlier, implies that
    \begin{equation}
        \frac{d^{n}}{dt^{n}} \Big[ D_M(\rho_t \rr \tfrac{\one}{d}) - D_N(\rho_t \rr \tfrac{\one}{d})  \Big] \Big|_{t=0} = 0,
    \end{equation}
    which implies, after evaluating the derivatives as above, that for all $n \geq 2$
    \begin{equation}
    \label{eqn:ordering-equality-derivative-condition}
        \textstyle\sum_i  \tr(M_i) \, (m_i)^{n} - \textstyle\sum_j  \tr(N_j) \, (n_j)^{n} = 0.
    \end{equation}
    To put this expression into a more useful form we define some notation for the normalized POVM elements of $M$ and $N$. 
    Let $\mathcal{M}=\{\mu \; | \; \mu = \frac{M_i}{\tr(M_i)}, \; M_i \in M\}$ 
    and
    $\mathcal{N}=\{\mu \; | \; \mu = \frac{N_j}{\tr(N_j)}, \; N_j \in N\}$
    be the sets of normalized POVM elements in each measurement. Let $\Gamma = \mathcal{M} \cup \mathcal{N}$ be their union, which contains all the normalized POVM elements from both measurements. And for any $\mu \in \Gamma$, let
    \begin{equation}
        V_M(\mu) = \sum_{M_i \in M(\mu)} \tr(M_i) ,
        \qquad \qquad
        M(\mu) = \{M_i \in M \; | \; \tfrac{M_i}{\tr(M_i)}=\mu  \},
    \end{equation}
    be the total volume of $\mu$ in $M$, and likewise for $V_N(\mu)$ in $N$.
    In these terms \eqref{eqn:ordering-equality-derivative-condition} becomes
    \begin{equation}
    \label{eqn:ordering-equality-volumes-sum}
        \sum_{\mu \in \Gamma} \; \big[ V_M(\mu) - V_N(\mu) \big] \, \tr(\mu X)^n = 0
    \end{equation}
    for all $n \geq 2$. Since this holds for all traceless Hermitian $X$, we can obtain that all $V_M(\mu)=V_N(\mu)$ by using a suitable $X$, as will be seen presently.

    Let $\Gbar = \Gamma \setminus \{\frac{\one}{d}\}$ be the set of all $\mu \in \Gamma$ that are not the maximally mixed state, and let $k=1,2,\ldots,K$ index the $\mu_k \in \Gbar$ (which is a finite set). Suppose there exists a traceless Hermitian matrix $\Xbar$ such that the inner products $r_k = \tr(\mu_k \Xbar)$ form a set of distinct (that is, nonrepeating) nonzero numbers, \ie~such that all $r_k \neq 0$ and such that $r_k-r_{k'}=0$ only if $k=k'$. Note that $\tr(\frac{\one}{d}\Xbar)=0$ since $\Xbar$ is traceless, so the maximally mixed case must be treated separately. We now assume such an $\Xbar$ exists, which will be shown later.
    
    If such an $\Xbar$ exists then \eqref{eqn:ordering-equality-volumes-sum} must hold for it, which yields that for all $n \geq 2$,
    \begin{equation}
    \label{eqn:ordering-equality-polynomial}
        \sum_{k=1}^{K} \; \big[ V_M(\mu_k) - V_N(\mu_k) \big] \, (r_k)^n = 0
    \end{equation}
    with $r_k=\tr(\mu_k \Xbar)$ a set of distinct nonzero numbers. (Note that any maximally mixed component of $\Gamma$ vanished from this equation since $\tr(\Xbar\one)=0$.) It is only possible to satisfy this equation for all $n \geq 2$ if the coefficients $V_M(\mu_k) - V_N(\mu_k)$ are zero, a consequence of the linear independence of polynomials/power series. To see this one can rewrite the above as a vector equation, where we take a linear combination with coefficient $c_k$ multiplying a vector $(r_k^2, r_k^3, \ldots, r_k^{L+1})^T$ for each $k$, so that we have a combination of $K$ vectors each of length $L$, where $L \geq 1$ can be arbitrarily high (corresponding to the arbitrarily high order of derivatives). For $L \geq K$ these vectors are all linearly independent from one another, since (making use of the Vandermonde determinant~\cite{horn1985book}) we have
    \begin{equation}
        \det
        \begin{pmatrix}
            r_1^2 & r_1^3 & \ldots & r_1^{K+1} \\
            \ldots & \ldots & \ldots & \ldots \\
            r_k^2 & r_k^3 & \ldots & r_k^{K+1} \\
            \ldots & \ldots & \ldots & \ldots \\
        \end{pmatrix}
        =
        \left( \prod_{k=1}^{K} r_k^2 \right) \left(\prod_{\substack{k',k''=1\\k'<k''}}^{K} (r_{k''} - r_{k'}) \right) \neq 0,
    \end{equation}
    which is nonzero since all $r_k$ are nonzero and distinct. Since the vectors are linearly independent, \eqref{eqn:ordering-equality-polynomial} can hold only if all coefficients are zero. Thus for all $\mu_k \in \Gbar$,
    \begin{equation}
    \label{eqn:ordering-equality-volumes-Gbar}
        V_M(\mu_k) = V_N(\mu_k).
    \end{equation}
    Finally, consider the case where $\mu_0=\one/d \in \Gamma$. Observe that since $\sum_i M_i = \sum_j N_j$ it follows that $V_M(\one/d) + \sum_{k} V_M(\mu_k) = V_N(\one/d) + \sum_{k} V_N(\mu_k)$, which using \eqref{eqn:ordering-equality-volumes-Gbar} implies $V_M(\one/d)=V_N(\one/d)$. Thus we have finally shown that for all $\mu \in \Gamma$, 
    \begin{equation}
    \label{eqn:ordering-equality-volumes-gamma}
        V_M(\mu) = V_N(\mu).
    \end{equation}
    This provides a very strong sense in which the $M,N$ POVMs are related: they each contain the same normalized elements with the same total weight, and in particular every element $M_i$ is proportional to some $N_j$. 

    The condition \eqref{eqn:ordering-equality-volumes-gamma} is the same as stochastic equivalence~\cite{martens1990nonideal}. For completeness we describe how it allows stochastic maps between $M$ and $N$ to be constructed. This is most easily done by defining an intermediary POVM $S=(S_l)_l$ with elements $S_l = \mu_l \, V(\mu_l)$, where $V(\mu) = V_M(\mu) = V_N(\mu)$ and~$l$~is taken to index the $\mu \in \Gamma$. Each $M_i$ corresponds to some particular $\mu_{l(i)} = M_i/\tr(M_i)$. We can therefore construct the stochastic map $\Lambda_{l|i} = \delta_{l,l(i)}$ so that $S_l = \sum_i \Lambda_{l|i} M_i$. In the other direction, we have $\Lambda_{i|l} = \frac{\tr(M_i)}{V(\mu_l)} \delta_{l,l(i)}$ so that $M_i = \sum_l \Lambda_{i|l} S_l$. Thus we have stochastic equivalence $M \leftrightarrow S$. The same construction can be obtained to find $N \leftrightarrow S$. So finally, composing the relevant maps, we have stochastic equivalence $M \leftrightarrow S \leftrightarrow N$. 

    It only remains to prove the existence of an $\Xbar$ of the type assumed above. In fact, almost all $X$ in the space $\XX$ of traceless Hermitian matrices fit the desired criteria. To see this we consider the complement of the allowed set. Consider the sets $\chi_k = \{X\in \XX \, | \, \tr(\mu_k X)=0 \}$. It cannot hold that $\chi_k = \XX$, because this would imply $\mu_k = \one/d$, which is false by assumption $\mu_k \in \Gbar$. But since $\chi_k$ is the set of solutions of a homogeneous linear equation, it is a linear subspace of $\XX$. Since it is not the whole space, it must be a linear subspace of strictly lower dimension (a hyperplane). Similarly, consider the sets $\chi_{k,k'} = \{X\in \XX \, | \, \tr((\mu_k - \mu_{k'}) X)=0 \}$ for all $k \neq k'$. Again $\chi_{k,k'}\neq \XX$ is not the whole space (this would imply $\mu_k = \mu_{k'}$), and is defined by a linear equation, and therefore must be a hyperplane. Each of these hyperplanes is a set of measure zero in $\XX$. Now consider the set $\chi$ which is the union of all these $\chi_k$ and all these $\chi_{k,k'}$. This is the union of a finite number of sets of measure zero, and so $\chi$ also has measure zero. Therefore $\chi$ has a nonempty complement. Any element of this complement is a valid choice of $\Xbar$. Therefore an $\Xbar$ of the desired type exists. In practice one can find a valid $\Xbar$ by choosing a traceless Hermitian matrix randomly, and checking numerically a finite number of linear conditions. 
    
    This completes the proof.    

    The proof given here continued to assume finite dimension $d$ and finitely indexed POVM outcome~sets. The assumption of finite dimension appeared when the maximally mixed state $\one/d$ was~used. However, this step was not necessary, and can be avoided by working directly with $S_M$ and evaluating derivatives at another point. (Evaluating at $\one/d$ and using $D_M$, as we have done, simplifies the argument and also clarifies why only the higher derivatives $n\geq 2$ are relevant to relating $M,N$.) But relaxing the $d<\infty$ assumption also makes it possible that $V(\mu)=\infty$, which introduces some additional complications. The assumption that the POVMs are at least discretely indexed is essential to the present form of the argument---there are only a discrete number of derivatives to work with in \eqref{eqn:ordering-equality-polynomial} for each $\mu$. However the relaxation from finite to countable outcome sets can be pursued, taking care to extend the linear independence of the polynomial constraints to infinite power series constraints. Extending the present results to infinite dimension and arbitrary POVM outcome sets will be a useful topic for continued investigation.
\end{proof}

\begin{figure*}
    \centerline{\includegraphics[width=.9\textwidth]{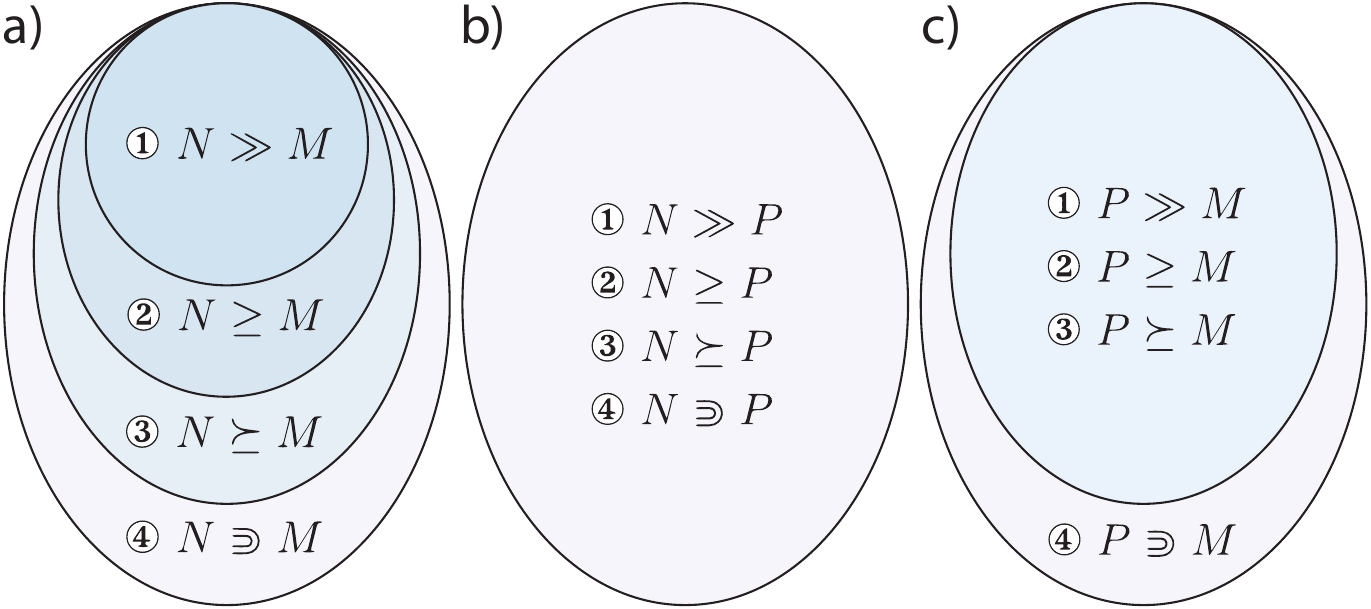}}
    \caption{
    Main results of the paper. Comparison of \circled{172} stochastic, \circled{173} relative entropy, \circled{174} entropy, and \circled{175} linear orderings for a) generic situation of two POVMs, b) more informative measurement is projective, c) less informative measurement is projective. Additionally, \circled{172}--\circled{174} induce the same equivalence classes (Theorem~\ref{thm:equality}).
    \label{fig:main}
    }
\end{figure*}

\section{Physical interpretation}

We summarize here the main results and interpret them physically. A general measurement is described by a positive operator-valued measure (POVM). Intuitively, some measurements can be more informative than others. There are different ways to make this intuition precise. We conceived four different mathematical definitions, which we interpret physically as follows.

The first is \circled{172} stochastic ordering, which means that the outcomes of the less informative measurement can be simulated by classical postprocessing (relabeling and re-binning outcomes, possibly with the help of a classical coin) of the finer one.

The second is order measured by \circled{173} relative entropy. Physically, if one measurement is more informative than the other in this sense, it means that for every pair of two states, it can see a larger difference between the statistics of measurement outcomes these states produce.

The third is order measured by \circled{174} observational entropy, which is the relative entropy between the state and the maximally mixed state. More informative measurement means that for any state, the observer sees larger differences between outcomes produced by this state and those produced by the maximally mixed state, than if they performed the less informative measurement.

Finally, the fourth order is given by \circled{175} linear relation, which is the hardest to interpret due to potentially negative factors. This indicates a situation in which the observer first collects the entire observed probability distribution and then allows to both add and subtract certain outcome probabilities to determine what would be the probability distribution observed when performing the less informative measurement if they chose to measure it. While less interpretable, this plays an important mathematical role connecting the others.

The main results of the paper are summarized in Fig.~\ref{fig:main}. The series of implications means that if the observer designed the situation via classical postprocessing, they will also see a larger difference between statistics of outcomes, a larger difference in entropies, and could determine the probability distribution of outcomes of the less informative measurement by adding and subtracting outcome probabilities with certain weights. If the more informative measurement is projective, all four situations are identical. If the less informative measurement is projective, the first three situations are identical. The fourth situation is relevant and differs from others only in situations where the more informative measurement is a non-projective POVM. It is also special in that only the fourth generates different equivalence classes from the other three.

\section{Concluding remarks}

In this article we investigated several senses in which measurements $M,N$ may be coarser, finer, or equivalent to one another. The result that the stochastic post-processing equivalence $N \equiv M$ is equivalent to both entropy equality $S_N=S_M$ and relative entropy equality $D_N=D_M$ (equalities implied for all states) lends to the conclusion that there is an essentially unique way for POVMs to be physically equivalent. Meanwhile, the results on partial ordering confirm that the stochastic post-processing partial order $N \gg M$ dominates the entropic orderings: coarseness in the stochastic sense implies coarseness in all other senses, but not necessarily vice versa. These results can both be interpreted in terms of monotones in the resource theory of measurements~\cite{guff2021resource}, and fit with the existing heuristics on measurement usefulness~\cite{martens1990nonideal,buscemi2005clean,haapasalo2017optimal,skrzypczyk2019robustness,leppajarvi2021postprocessing}.

We note several avenues of interest for continued study. Most directly, relaxing the assumptions of finite dimension and discrete outcome sets in the present results will be an important generalizing step. One can also compare the quantum results obtained here to an analogous classical case. Additionally, in light of the fact that $S_N \geq S_N$ requires a linear relation $N \Supset M$, an interesting question is whether meaningful algebraic constraints on the matrix $\alpha_{ji}$ can be identified that are sufficient to ensure the entropy (or relative entropy) inequality. Identification of such constraints could shed light on the nature of entropy-coarser measurements at the POVM element level. Lastly, as mentioned in the introduction, another essential extension is to study the orderings obtained by replacing $D \to D^\alpha$ by the $\alpha$-Renyi divergences. In particular the $D^{\max}$ ordering is conjectured to collapse the hierarchy, analogous to \eqref{eqn:guessing-probs} and the results of~\cite{buscemi2018reverse,buscemi2015degradable,skrzypczyk2019robustness}. 

Beyond illuminating the structure of the space of POVMs, we expect the results to be of use as a mathematical tool in ongoing studies of observational entropy in thermodynamics~\cite{safranek2021brief,strasberg2020first}, where characterizing coarser/finer measurements can bound coarse-grained entropies in physical systems, and these results can be a useful tool for proving information theoretic properties of statistical mechanical entropies.

\begin{acknowledgements}

The authors thank Francesco Buscemi and Andreas Winter for helpful discussions. J.S. acknowledges support by MICIN with funding from European Union NextGenerationEU (PRTR-C17.I1) and by Generalitat de Catalunya. J.S. also acknowledges support from the Beatriu de Pinós postdoctoral fellowship under the Catalan Ministry of Research and Universities and  EU Horizon 2020 MSCA grant 801370.
A. T.-B. received the support of a fellowship from ``la Caixa” Foundation (ID 100010434, with fellowship code LCF/BQ/EU21/11890119). Research at Perimeter Institute is supported in part by the Government of Canada through the Department of Innovation, Science and Economic Development Canada and by the Province of Ontario through the Ministry of Colleges and Universities. 
D.\v{S}.~acknowledges the support from the Institute for Basic Science in Korea (IBS-R024-D1). 
\end{acknowledgements}


\bibliography{biblio}

\clearpage

\end{document}